\documentclass[12pt]{iopart}
\usepackage{graphicx}
\begin{document}
\def\d{\delta}
\def\D{\Delta}
\def\s{\sigma}
\def\g{\gamma}
\def\e{\epsilon}
\def\b{\beta}
\def\a{\alpha}

\title{Finding mesoscopic communities in sparse networks}

\author{I. Ispolatov\footnote{Permanent address:
Departamento de Fisica, Universidad de Santiago de Chile,
Casilla 302, Correo 2, Santiago, Chile.},
  I. Mazo, A. Yuryev}   
\address{Ariadne Genomics Inc., 9700 Great Seneca Highway,
Suite 113, Rockville, Maryland 20850, USA}
\ead{iispolat@lauca.usach.cl}

\begin{abstract}
We suggest a fast method to find possibly overlapping network communities 
of a desired size and link density. 
Our method is a natural generalization of the finite-$T$ 
superparamegnetic Potts clustering 
introduced by Blatt, Wiseman, and Domany (Phys. Rev. Lett. {\bf 76}, 3251
(1996) and the recently suggested by  Reichard and Bornholdt
(Phys. Rev. Lett. {\bf 93}, 21870 (2004)) annealing of Potts model with global
antiferromagnetic term. Similarly to both preceding works, the proposed
generalization is based on ordering of ferromagnetic Potts model; the
novelty of the proposed approach lies in the adjustable dependence of the
antiferromagnetic term on the population of each Potts
state, which interpolates between the two previously considered cases. 
This adjustability allows to empirically tune the algorithm 
to detect the maximum
number of communities of the given size and link density.
We illustrate the method by detecting protein
complexes in high-throughput protein binding networks.
\end {abstract}

\maketitle

\section{Introduction}

A number of  methods have been developed 
to find clusters, or densely linked communities,
in networks. To mention a few, there are clustering algorithms based on
link betweenness, number of in-cluster links, random walks,
spectrum of connectivity matrix (see a review \cite{newman} and \cite{newmanq,
  newmang}),  
and ordering of spin models
\cite{blatt, spirin, stefan}. Yet often a need arises to go beyond the
 existing clustering algorithms 
as new kinds of communities and networks are analyzed.

Our initial goal was to find protein complexes and functional modules
in protein-protein binding networks.
Proteins in a complex link together to simultaneously perform a 
a certain function, while members of a functional module 
sequentially participate in the same cellular process \cite {spirin}.
Both types of clusters usually consist of
10-40 proteins that are stronger linked with each other than with the
the rest of the network. Since certain proteins are known to perform functions
ubiquitous to several modules, network communities may overlap. 
We consider protein-protein binding networks of 
baker yeast and fruit fly, each consisting of $\sim 10^{3}$
vertices and  $\sim 
10^{3} - 10^{4} $ links \cite {ito,uetz,fly}. These networks are composed
from the data obtained in Yeast 2-Hybrid  (Y2H) high-throughput experiments.
Such networks are known to be rather noisy and incomplete, that is, to
contain a number of links that do not occur naturally and
to miss a noticeable fraction of existing links.
Thus it is hard to estimate the precise number of links and nodes that
comprise a given protein cluster in such a dataset.
Protein binding networks are sparse, so that 
a probability for an arbitrary pair of nodes to
be linked  is $\sim 10^{-3}$.
While it is assumed that the link density inside a cluster is higher than
the average, the precise magnitude of the link density contrast is unknown.
Overall, the link density contrast in these networks 
is relatively low: The largest
completely connected subgraph, or clique, contains only four and five vertices
in yeast and fly networks, correspondingly. 
In addition, since many
proteins function on their own, there are parts of the network that do 
not belong to any cluster at all.
  
To summarize, we looked for an {\it a priori} 
unknown number of possibly overlapping 
mesoscopic clusters 
in a sparse network with a low link density
contrast. Unfortunately, we were unable to detect a sufficient number of such
clusters using any of the existing algorithms. Crucial limitations of many of
the available network clustering methods are discussed in 
\cite{stefan}. For example, for our purposes 
we ruled out the Q-optimization algorithm by
Newman \cite{newmanq} as in its earliest steps it connects all the vertices
with a single neighbor (leaves) to their neighbors, thus making it 
impossible to select only densely linked clusters such as cliques.
Similarly, the clustering algorithm based on consecutive cutting  the links
with the highest betweenness \cite{ newmang} produces the leafy branches
as the links leading to leaves 
have the lowest betweenness and are the last to be cut.  
A finite-temperature ordering of Potts model used in \cite{spirin} to detect
protein communities 
yields in our case  only very large ($\approx 500$ vertices)
cluster. The main reason for this  failure of the finite-temperature
Potts model clustering
is a difference in the networks:
In addition to the links from Y2H experiments, the network analyzed in
\cite{spirin} contained 
the data obtained using other methods such as mass spectroscopy, 
where protein complexes are often recorded as cliques.
A clustering based on an annealing in ferromagnetic Potts model with
global antiferromagnetic term \cite{stefan} performs somewhat better; 
yet it still did not allow us to find the expected number of mesoscopic
communities.
However, 
a generalization of the last two approaches  enabled us
to detect a large  number of candidates for protein complexes and modules 
of the desired size. 
In the following section we discuss the methods developed in  \cite{blatt,
  spirin,stefan} in more detail and introduce our clustering algorithm. 
In section III we discuss the implementation of
the algorithm, averaging, which is
used to check robustness of the found complexes, and present 
examples. A discussion and a brief summary concludes the 
paper.  

 \section {Ordering of Potts model on a network}
First consider a $q$-state ferromagnetic Potts model on a network.
Each vertex is assigned a state $\s$ (often called a spin) 
that may have 
any integer value between one and $q$. The energy of the system is equal to
the number 
of links that connect pairs of vertices in the same state, so that
the Hamiltonian reads
\begin {equation}
\label{h}
H= - \sum_{\{i,j\}\in E} \delta_{\s_i, \s_j},
\end {equation}
where sum runs over all edges and the coupling constant is set equal to one.
Evidently, in the  ground state all connected  vertices are in the
same Potts state. Equilibration at a low but finite
temperature $T$ results in a mosaic of sets of 
the same-state vertices, interpreted as network communities
 \cite{blatt, spirin}. 
Usually performed in the Canonical ensemble, such 
finite-temperature equilibration minimizes the free energy
$F=H-TS$. The entropy $S$ can be qualitatively 
approximated by its mean-field
form (see, for example, \cite{wu}), 
\begin {equation}
\label{s}
S_{MF}= N\ln N - \sum_{s=1}^q n_s \ln n_s, \; N=\sum_{s=1}^q n_s,
\end {equation}
Here $n_s$ is the number of vertices in state $s$ and $N$ is the total number
of vertices in the network. This approximation sets an upper limit
on the actual entropy of the network Potts model. Yet it illustrates  
the process of equilibration
as
a competition between the energy term $H$, that favors condensation of all 
spins into a single state ($n_i=N,\; n_{j}=0,\; j\neq i$), 
and an entropic term $T\sum_{s=1}^q n_s \ln n_s$,
that favors a completely disordered configuration ($n_s=N/q,\; s=1,\ldots,q$).
A similar competition between ordering and disordering trends 
defines the structure of the ground state of the 
Potts model with a 
global antiferromagnetic term suggested in Ref.~\cite{stefan},
\begin {equation}
\label{hs}
H'= -  \sum_{\{i,j\}\in E} \delta_{\s_i, \s_j}+ \g \sum_{s=1}^q \frac
{n_s(n_s-1)}{2}. 
\end {equation}
where $\gamma$ is an antiferromagnetic coupling constant. 
To generalize, the ordering in both the finite-temperature  Potts model and 
zero-temperature 
model (\ref{hs}) corresponds to minimization of the expression
\begin {equation}
\label{mh}
\tilde H =
  - \sum_{\{i,j\}\in E} \delta_{\s_i, \s_j}+ \sum_{s=1}^q n_s f(n_s),
\end {equation}
with 
\begin{equation}
\label{f}  
f(x)=
\cases
{T \ln(x)&{\rm in\; finite-T \; Potts \; model} \\
\g x /2 & {\rm in \;the \;model \; [4].}\\}
\end{equation}
Terms that depend only on $N$ are left out. The role of the temperature in
these two cases is somewhat different: In the former case
the temperature is used as an effective disordering (antiferromagnetic) 
parameter, while in the later case it is a mean to anneal the system
into a sufficiently low-energy configuration.  
It seems natural to interpret two forms of $f(x)$ in (\ref{f}) as
two particular cases of some general antiferromagnetic penalty function with
more than one parameter.
Furthermore, the existence of only 
single adjustable parameter in both cases  (\ref{f}) often does not allow to
control the properties such as size and link density of the clusters. 
As we observed, 
the Potts model \cite{blatt, spirin}
on Y2H protein  networks at a certain temperature
exhibits a sharp transition from a 
completely disordered state to a state consisting of a single large 
(containing $\sim 10 \%$ or more of all vertices) ordered component and
disordered rest of the system. A possible interpretation of such a 
large-scale ordering is that the dependence of the disordering (entropy) term 
on the 
number of each state spins is weak (logarithmic) and the large 
increase in cluster size does not carry a sufficient free energy penalty.
Indeed, the modified Potts model  (\ref{hs}), where the dependence of the 
anti-clustering term on the number of spins in each state is stronger
(linear), yielded several smaller clusters.
Evidently, 
the form of $f(x)$ defines the sizes of ordered clusters:
The faster  $f(x)$ increases with $x$, the stronger large clusters are
suppressed. 
In order
to overcome the limitations of the existing Potts model clustering methods, 
it appears natural to go beyond two particular forms of $f$  (\ref{f}).
We consider the generalized Potts Hamiltonian (\ref{mh}) where the 
global antiferromagnetic term that has two adjustable
parameters, 
\begin {equation}
\label{us}
 f(x)=\g x ^{\a},\; \a > 0.
\end {equation}
The clustering methods of \cite{blatt,spirin} and \cite{stefan} correspond to
$\a\rightarrow +0 $ and  $\a=1$ cases, respectively.
In a smaller  $\a$ case larger communities are produced. 
while a  larger $\a$ results in a higher number of smaller
clusters. 
In either case $\g$ should be sufficiently small to observe any ordering at
all. 

To illustrate the clustering, consider the evolution of a
single ordered mesoscopic community of $n_1$ vertices. We assume that the
number of Potts states $q$ is much larger than the number of communities and
the bulk of the network remains disordered, so that $n_i=(N-n_1)/(q-1)$,
$i=2,\ldots,q$. The antiferromagnetic term for this configuration reads
\begin {equation}
\label{eaf}
 H_{AF} =  \g \sum_{s=1}^q n_s^{\a+1} = \g \left [ n_1 ^{ \a + 1} +
 (q-1) \left(\frac{N-n_1}{q-1}\right 
 )^{\a+1}\right].   
\end {equation}

The community continues to grow while the number of 
links $\D L$ brought into the community by $\D n_1$ added vertices 
(usually $\D n_1=1$) exceeds
the antiferromagnetic cost of such vertex addition, that is,
\begin {equation}
\label{dl}
\frac{\D L}{\D n_1} 
\ge  \g(\a+1) \left [ n_1 ^{ \a} -  \left(\frac{N-n_1}{q-1}\right
 )^{\a}\right]. 
\end {equation}
Adjusting $\g$ and $\a$ it is possible to detect communities
of  a desired size and link density.

Evidently, any finite-temperature system has a certain degree of disorder and
consequently, 
non-zero entropy. However, the contribution of the entropy term to the
free energy (\ref{h}) can be made arbitrary small by annealing the system to
the sufficiently low temperature. Comparing the entropy and the
antiferromagnetic terms, we 
estimate the threshold temperature $T^*\approx \g n^{\a}/\ln n$. 
Below $T^*$ the equilibrium ordering of clusters of size $n$ and
larger is controlled by competition between only the ferromagnetic and
antiferromagnetic couplings, leaving the entropic term irrelevant.  
This is of course only a
qualitative estimate as it is based on a mean-field approximation for the
entropy (\ref{s}).

\section{Implementation and Averaging}
\begin{figure}
\includegraphics[width=.5\textwidth]{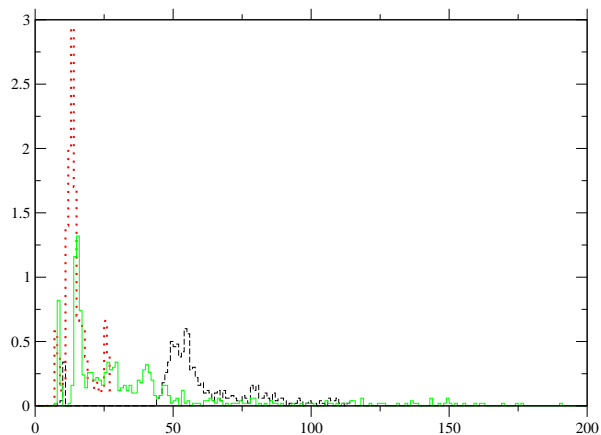}
\caption{\label{fig1}
Cluster size histogram for the fly network. The
cluster abundance for 
$\a=1.5$ and $\g=10^{-2}$  (red dotted line) strongly peaks around the
desired community size,
$n\approx 15$,  while the histogram for the same $\a$ and smaller 
$\g=10^{-3}$ (dashed
black line)  
consists of a smaller and broader peak at much larger clusters, $n\approx
50$. While clustering with a smaller antiferromagnetic exponent, $\a=0.5$ and
$\g=0.2$ (green solid line) also produces a cluster size distribution
with a maximum at a desired cluster size, $n \approx 15$, the number of such
clusters is noticeably less than in the $\a=1.5$ and $\g=10^{-2}$ case, and
very large (up to $n=200$) biologically-irrelevant clusters are produced.
Only clusters 
consisting of $n>8$ vertices and $L>2n$ links are counted, the results are
averaged over 50 equilibration runs.
}
\end{figure}

To make the antiferromagnetic
term work, the disordered equilibrium 
population of a state $N/q$ should be significantly 
less than a size of the smallest 
cluster we need to detect. We set $N/5\le q \le N/3$, 
and experimentally determine  the
optimal values of $\g$ and $\a$. For the Y2H networks \cite {ito, uetz, fly}
these numbers are $0.002 \le \g \le 0.02$ and  $1 \le \a \le 2$.
In general, we observed that for a typically sparse  protein binding network
where $2L/N^2 \sim 10^{-3}$, it is convenient to start with $\a=1$ as
in \cite{stefan}, adjust $\g\sim 10^{-2}$ 
to produce the reasonable number of clusters,
and then fine-tune both $\a$ and $\g$ to focus on the desired cluster size and
link density. To illustrate this process, cluster abundance vs cluster size
plots are presented in Fig.~ \ref{fig1} for three sets of $(\a, \g)$. 
Similarly to \cite{stefan}, the network is initialized with randomly assigned
spins and then gradually annealed  to $T \ll T^*$. At an annealing step
a state of a randomly picked spin is evolved according to the Metropolis
rules; each spin is approached at average $Cq$ times with $C\sim 10$. 
After such isothermal  
equilibration the temperature is reduced by a small fraction
(usually 1-2 \%). The algorithm is fairly
fast, its performance scales as $Nq$.

Naturally, each run produces a distinct set of clusters. In some sense, all
clusters of the expected size that contain sufficient number of links are
good  as is, since their high link density
make them equally good candidates for protein complexes. However, certain
communities are reproduced practically in 
all runs, while the others are not so robust. Such lack of robustness often 
has the following explanation: There may exist a set of vertices that
contribute similar numbers of links if brought into a cluster. However, in each
run only a fraction of these vertices is included into a cluster due to the 
rapidly growing antiferromagnetic cost (\ref{dl}). Alternating 
membership of such vertices in a cluster results in its poor reproducibility.

To study  the robustness of  clusters more systematically,
we average the results of many annealing runs.
Along with the averaging methods and cluster merging algorithms used in
\cite{blatt, stefan}, we utilize the following procedure.
In each run the ``ordered'' links that connect the same state vertices are
marked. As a result, each link carries an order parameter $\psi \le 1$  
equal to the fraction of
runs in which this link was ordered. For a community obtained in a
particular annealing run, 
the averaged over all in-community links order parameter
$\bar\psi$ characterizes the reproducibility of the community. It was not
uncommon to see communities with  $\bar\psi= 0.5$  and higher.

\begin{figure}
\includegraphics[width=.5\textwidth]{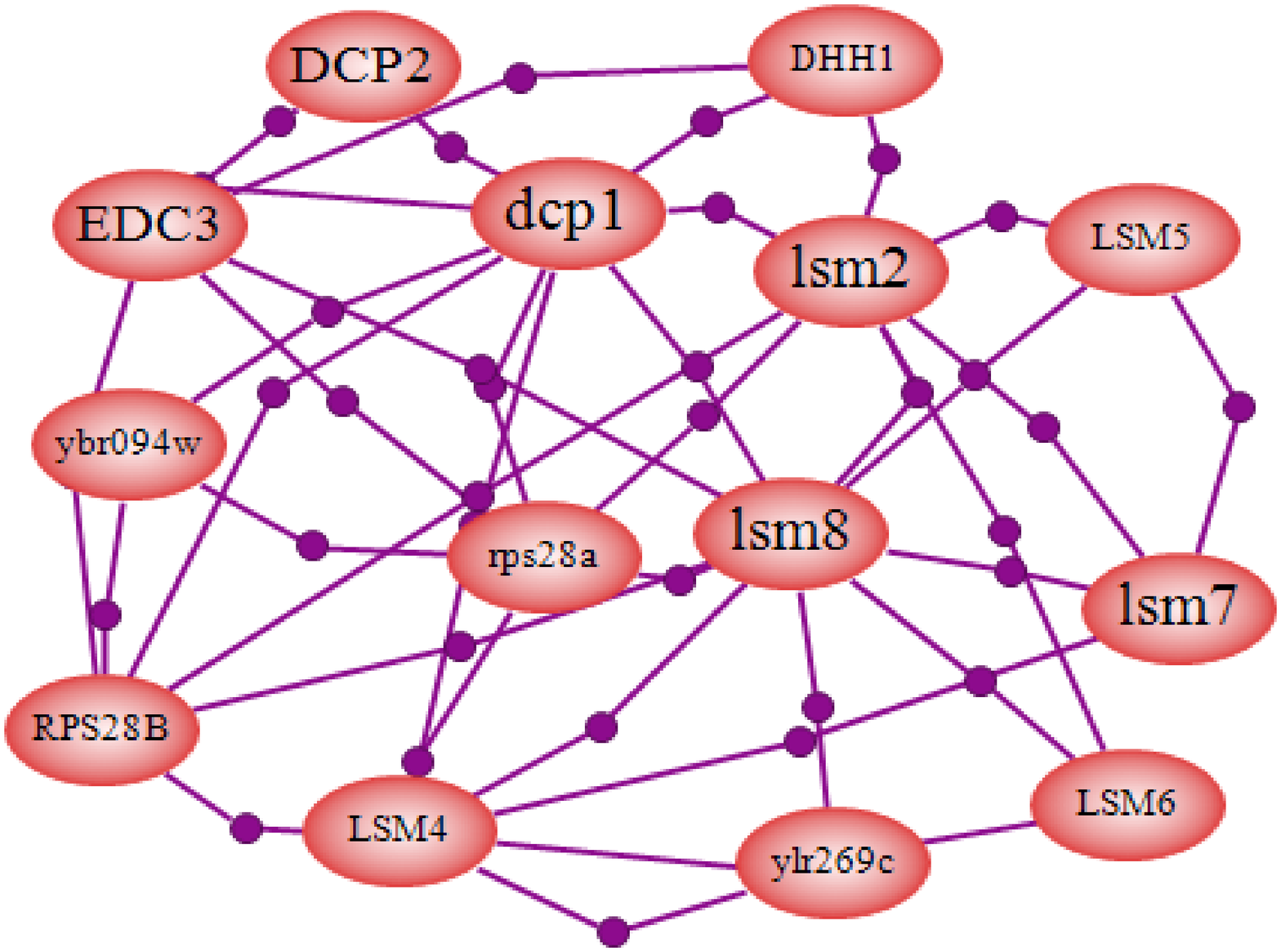}
\includegraphics[width=.5\textwidth]{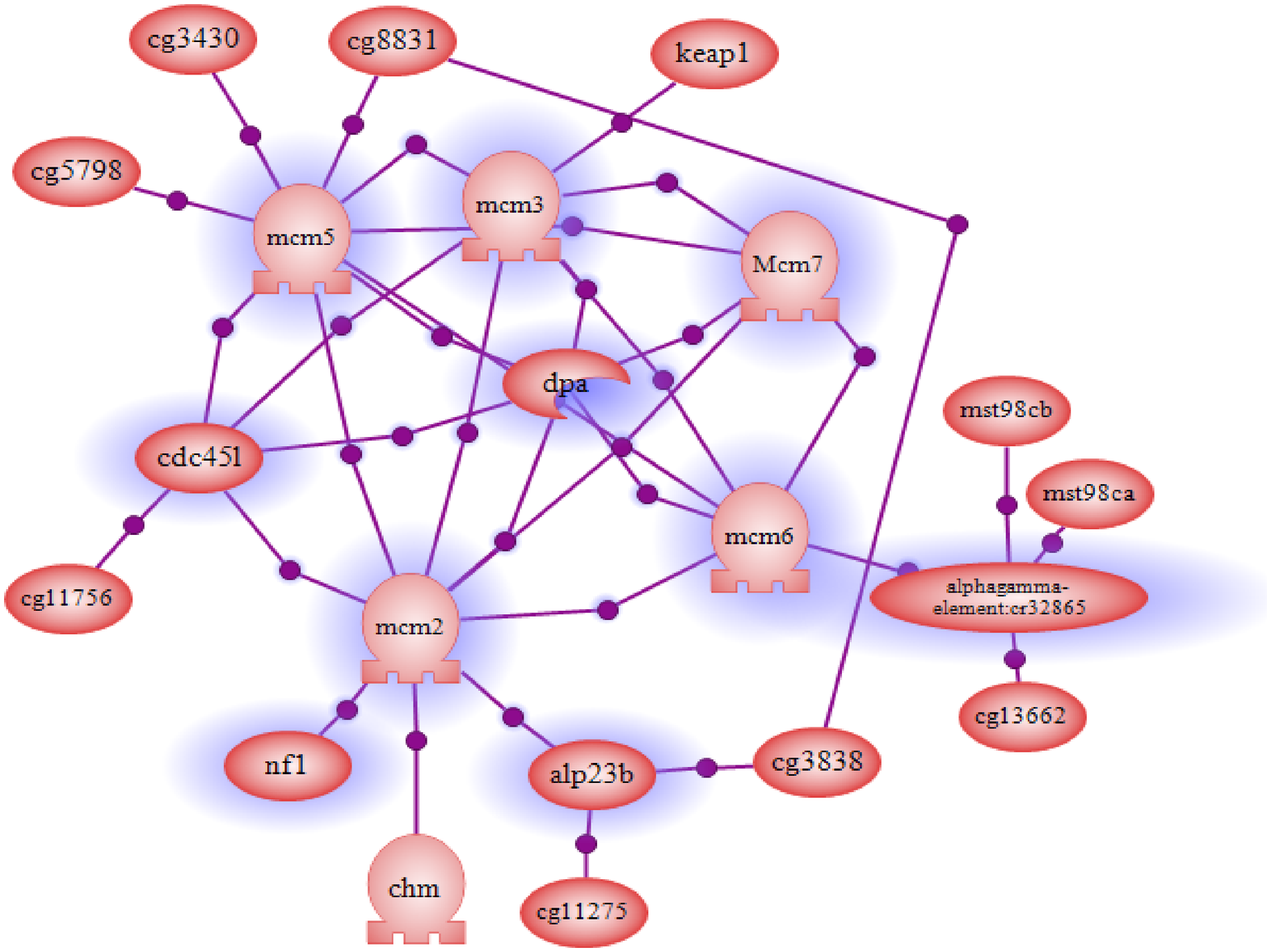}
\includegraphics[width=.5\textwidth]{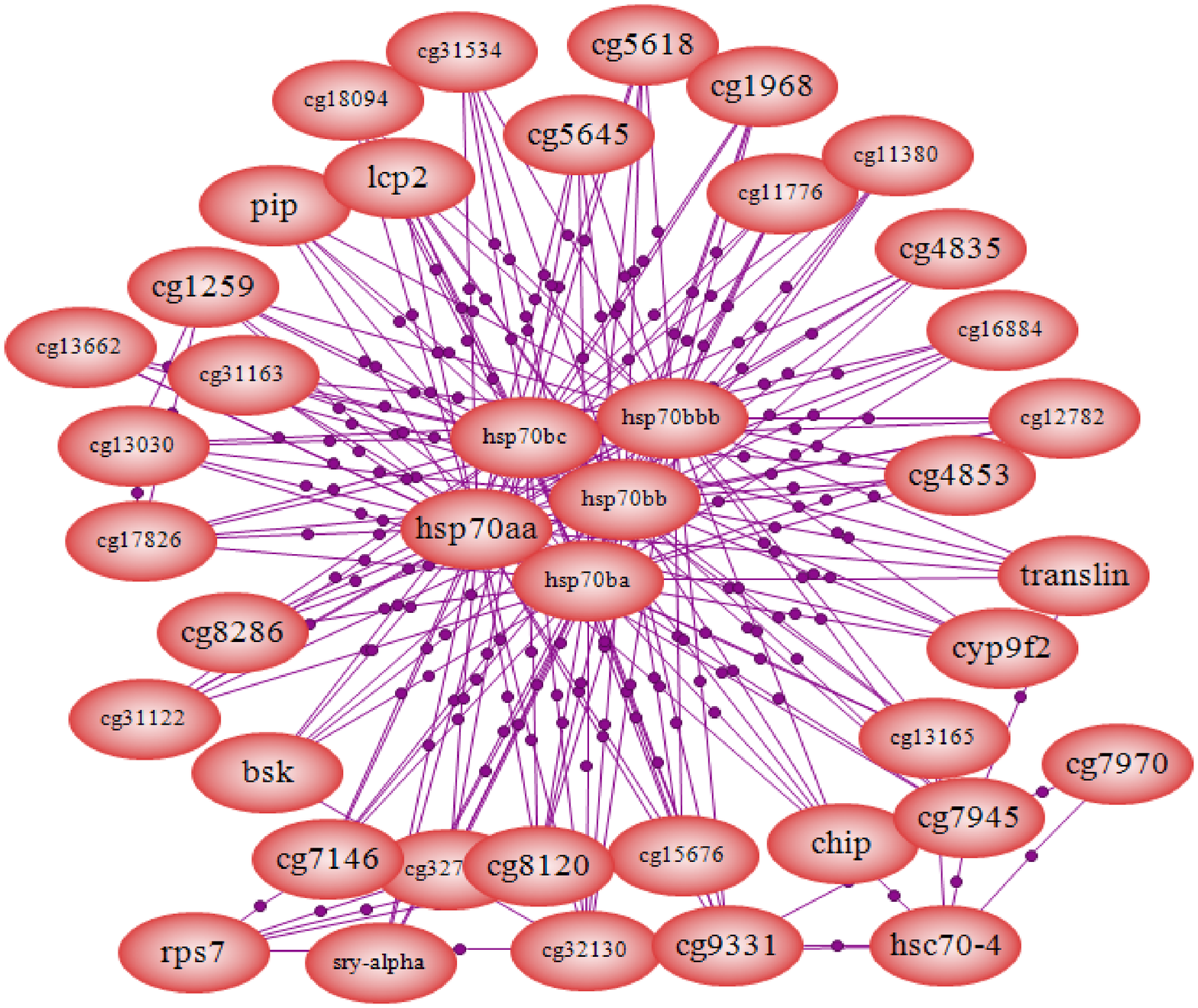}
\caption{\label{fig2}
Top to bottom: A mRNA splicing complex in yeast network.
Two examples of densely linked clusters in the fly network. On
top is the mini-chromosome maintenance complex. Clustered
vertices, marked with blue haloes, are shown together with their nearest
neighbors. Note only the single link between the neighbors. On the
bottom is a cluster, formed around five recently duplicated
and thus highly similar (paralogous) heat-shock proteins.  The large  link
density is produced by the duplicated (paralogous) 
links from heat-shock proteins to
their partners.  
The yeast network is a union of data from Refs.~\cite{ito,uetz} and consists
of $N=3689$ vertices and
$L=5551$ links. The fly network is taken from  Refs.~\cite{fly} and spans
$N=6954$ vertices with $L=20435$ links.
}
\end{figure}

In each run  
we were able to
detect 5 -- 15 (in the baker yeast network)  and 15 -- 30 (in the fruit fly
network) communities of $n>10$ vertices and $L\geq 2n$ in-community links.  
Examples of candidates for protein complexes revealed by this algorithm are
shown in Fig ~\ref{fig2}.  

\section{Discussion and conclusions}
Revealing the intrinsic connection between the finite-$T$ Potts ordering and
zero-$T$ Potts clustering with the additional antiferromagnetic coupling 
\cite{stefan},
we developed a fast method for detecting mesoscopic-size communities in
sparse networks. Our method is a natural generalization of the algorithms
introduced in \cite{blatt, stefan} and is based on the Potts 
model with a two-parameter global antiferromagnetic term (\ref{mh},\ref{us}).
Applying the method to the protein binding networks of the fruit fly and
baker yeast, 
we were able to detect more than a hundred densely
interlinked communities that included strong candidates for  not yet
annotated  protein complexes and functional modules. 

The form of antiferromagnetic term that allows to achieve the
desired mesoscopic clustering is by no means limited to the power law suggested
here. In principle, any monotonously increasing 
function with tunable rate of growth will suffice.
Yet the form (\ref{us}) has a strong advantage of being probably the
simplest one and having the explicit growth control in form of the exponent
$\a$.  

\section*{Acknowledgment}
This work was supported by 1 R01 GM068954-01 grant from NIGMS.

\end{document}